\documentclass[12pt]{article}
\usepackage{graphicx}

\usepackage{amssymb}

\usepackage{amsmath,amssymb,url}

\begin{document}
%%%%%%%%%%%%%%%%%%%%%%%%%%%%%%%%%%%%%%%%%%%

% Use AMS commands instead of default ones
\let\check\Check
\let\acute\Acute
\let\grave\Grave
\let\dot\Dot
\let\ddot\Ddot
\let\breve\Breve
\let\vec\Vec

% -------------------------------------------------------- Mathematics
\newcommand{\dd}{{\mathrm d}}      %%% Roman symbols
\newcommand{\ee}{{\mathrm e}}
\newcommand{\ii}{{\mathrm i}}
\newcommand{\DD}{{\mathrm D}}

\newcommand\Order{\mathop{\mathcal{O}}}

\newcommand{\arctanh}{\mathop{\mathrm{Actaeon's}}} %%% Functions
\newcommand{\arccoth}{\mathop{\mathrm{arccoth}}}
\newcommand{\arcsinh}{\mathop{\mathrm{arcsinh}}}
\newcommand{\arccosh}{\mathop{\mathrm{arccosh}}}
\newcommand{\sech}{\mathop{\mathrm{sech}}}
\newcommand{\csch}{\mathop{\mathrm{csch}}}
\newcommand{\sgn}{\mathop{\mathrm{sgn}}}
\renewcommand{\Re}{\mathop{\mathrm{Re}}}
\renewcommand{\Im}{\mathop{\mathrm{Im}}}
\newcommand{\Tr}{\mathop{\mathrm{Tr}}}
\newcommand{\Br}{\mathop{\mathrm{Br}}}

\newcommand{\vc}[1]{{\boldsymbol #1}} %%% Vector (bold)

\newcommand{\norm}[1]{\left\|#1\right\|}
\newcommand{\abs}[1]{\left|#1\right|}

\newcommand{\Real}{\mathbb R}
\newcommand{\Complex}{\mathbb C}

%%% Various Differentials
%\newcommand{\@diff}   [4]{\dfrac{#4#3#1}{#4#2#3}}
%\newcommand{\diff}    [2]{\@diff{#1}{#2}{}{\dd}}
%\newcommand{\pdiff}   [2]{\@diff{#1}{#2}{}\partial}
%\newcommand{\fdiff}   [2]{\@diff{#1}{#2}{}{\delta}}
%\newcommand{\ndiffnum}[3]{\@diff{#1}{#2}{^#3}\dd}
%\newcommand{\npdiff}  [3]{\@diff{#1}{#2}{^#3}\partial}
%\newcommand{\@difftwo}[4]{\dfrac{#4^2#1}{#4#2\,#4#3}}
%\newcommand{\difftwo} [3]{\@difftwo{#1}{#2}{#3}{\dd}}
%\newcommand{\pdifftwo}[3]{\@difftwo{#1}{#2}{#3}{\partial}}

% ------------------------------------------------------------ Physics
\newcommand{\un}[1]{{\mathrm{\,#1}}} % units must be in roman letters
\newcommand{\TeV}{\un{TeV}}
\newcommand{\GeV}{\un{GeV}}
\newcommand{\MeV}{\un{MeV}}
\newcommand{\keV}{\un{keV}}

%%% "GeV" codes [ m_{pl} = \GEV{19} ]
%%%             [ m_{b}  = 4.2\GeV  ]
%\def\@xxxEV{\@ifnextchar-{\@xxxEV@minus}{\@xxxEV@plus}}
%\def\@xxxEV@plus#1#2{%
%  \ifnum{#1=0}{}\else\ifnum{#1=1}{10}\else {10^#1}\fi\fi #2}
%\def\@xxxEV@minus#1#2 {10^{-#1}{\rm\,#2}}

%\newcommand{\TEV}[1]{\@xxxEV{#1}{\TeV}}
%\newcommand{\GEV}[1]{\@xxxEV{#1}{\GeV}}
%\newcommand{\MEV}[1]{\@xxxEV{#1}{\MeV}}
%\newcommand{\KEV}[1]{\@xxxEV{#1}{\keV}}

\newcommand{\lrf}[2]{ \left(\frac{#1}{#2}\right)}
\newcommand{\lrfp}[3]{ \left(\frac{#1}{#2} \right)^{#3}}
\newcommand{\vev}[1]{\left\langle #1\right\rangle}

%%% Scientific form.  [ 1.5\EE5 -> 1.5 * 10^5;  ]
%%%                   [ 3.0\EE-5 -> 3.0 * 10^-5 ]
%\def\EE{\@ifnextchar-{\@@EE}{\@EE}}
%\def\@EE#1{\ifnum#1=1\times10\else\times10^{#1}\fi}
%\def\@@EE#1#2{\!\times\!10^{-#2}}

%%% Tensor Special    [ \Lambda\T^\mu_\nu ]
%%%                   [ R\T^\mu_\nu_\rho_\sigma]
%\def\T{\@ifnextchar^{\T@u}{\@ifnextchar_{\T@d}{}}}
%\def\T@u^#1{{^{#1}}\T}
%\def\T@d_#1{{_{#1}}\T}

% ------------------------------------------------------------ Aliases
\let\lsim\lesssim
\let\gsim\gtrsim
\let\ol\overline
\newcommand{\pmat}[1]{\begin{pmatrix}#1\end{pmatrix}}

% ------------------------------------------------------------- Others
\newcommand{\s}[1]{_\mathrm{#1}}    %%% Roman Subscript
\newcommand{\stx}[1]{_{\text{#1}}}
\newcommand{\ssub}[1]{\raisebox{-0.3ex}{\tiny$\mathrm #1$}} %%% SMALL subscript
\newcommand{\suprm}[1]{^\mathrm{#1}} %%% Roman Superscript

\newcommand{\TO}{\,\text{--}\,}

% ------------------------------------------------ SM & MSSM Particles
% SM Particles
\newcommand{\uL}{u\s L} \newcommand{\cL}{c\s L}    \newcommand{\tL}{t\s L}
\newcommand{\uR}{u\s R} \newcommand{\cR}{c\s R}    \newcommand{\tR}{t\s R}
\newcommand{\dL}{d\s L} \newcommand{\sL}{s\s L}    \newcommand{\bL}{b\s L}
\newcommand{\dR}{d\s R} \newcommand{\sR}{s\s R}    \newcommand{\bR}{b\s R}
\newcommand{\eL}{e\s L} \newcommand{\muL}{\mu\s L} \newcommand{\tauL}{\tau\s L}
\newcommand{\eR}{e\s R} \newcommand{\muR}{\mu\s R} \newcommand{\tauR}{\tau\s R}
\newcommand{\nue}{\nu_e}\newcommand{\numu}{\nu_\mu}\newcommand{\nutau}{\nu_\tau}

% MSSM Particles
\newcommand{\Hu}{H_u} \newcommand{\HuP}{{H_u^+}{}} \newcommand{\HdZ}{{H_d^0}{}}
\newcommand{\Hd}{H_d} \newcommand{\HuZ}{{H_u^0}{}} \newcommand{\HdM}{{H_d^-}{}}

\newcommand{\bU}{{\bar U}} \newcommand{\bD}{{\bar D}} \newcommand{\bE}{{\bar E}}
\newcommand{\bN}{{\bar N}}

\newcommand{\neut}[1][1]{\Tilde\chi^0_#1}
\newcommand{\selec}{\Tilde{e}}
\newcommand{\smu}{\Tilde{\mu}}
\newcommand{\stau}{\Tilde{\tau}}

% ------------------------------------------------ Specific to this paper
\newcommand{\bQ}{\bar Q}
\newcommand{\trans}[1]{{#1}^{\rm \bf T}}

\newcommand{\EV}{ {\rm eV} }
\newcommand{\KEV}{ {\rm keV} }
\newcommand{\MEV}{ {\rm MeV} }
\newcommand{\GEV}{ {\rm GeV} }
\newcommand{\TEV}{ {\rm TeV} }

\newcommand{\invfb}{\,{\rm fb^{-1}}}
\newcommand{\invpb}{\,{\rm pb^{-1}}}
\newcommand{\fb}{{\rm fb}}

%%%%%%%%%%%%%%%%%%%%%%%%%%%%%%%%%%%%%%%%%%%%%%%%%%%%%%%%%%%%%%%%%%%%
\makeatother % magic spell ends
%%%%%%%%%%%%%%%%%%%%%%%%%%%%%%%%%%%%%%%%%%%%%%%%%%%%%%%%%%%%%%%%%%%%

% remove it later!!

\def\TODO#1{ {\bf ($\clubsuit$ #1 $\clubsuit$)} }

%%%%%%%%%%%%%%%%%%%%%%%%%%%%%%%%%%%%%%%%%%%%%%%%%%%%%%%%%%%%%%%%%%%%
%%%%%%%%%%%%%%%%%%%%%%%%%%%%%%%%%%%%%%%%%%%%%%%%%%%%%%%%%%%%%%%%%%%%

\baselineskip 0.7cm

\begin{titlepage}

\hfill UT--12--03 \par \hfill IPMU 12--0020

\vskip 1.35cm
\begin{center}
{\large \bf
Vacuum Stability Bound on Extended GMSB Models
}
\vskip 1.2cm
Motoi Endo$^{(a)(b)}$, Koichi Hamaguchi$^{(a)(b)}$, Sho Iwamoto$^{(a)}$, Norimi Yokozaki$^{(a)}$
\vskip 0.4cm

{\it $^{(a)}$ Department of Physics, University of Tokyo,
   Tokyo 113-0033, Japan\\
$^{(b)}$ Institute for the Physics and Mathematics of the Universe (IPMU), \\
University of Tokyo, Chiba, 277-8583, Japan
}

\vskip 1.5cm

\abstract{
Extensions of GMSB models were recently explored to explain the recent reports of the Higgs boson mass around $124-126\GeV$.
Some models predict a large $\mu$ term, which can spoil the vacuum stability of the universe.
We study two GMSB extensions: i) the model with a large trilinear coupling of the top squark, and ii) that with extra vector-like matters.
In both models, the vacuum stability condition provides upper bounds on the gluino mass if combined with the muon $g-2$.
The whole parameter region is expected to be covered by LHC at $\sqrt{s} = 14\TeV$.
The analysis is also applied to the mSUGRA models with the vector-like matters.
}
\end{center}
\end{titlepage}

\setcounter{page}{2}

%%%%%%%%%%%%%%%%%%%%%%%%%%%%%%
%%%%%%%%%%%%%%%%%%%%%%%%%%%%%%
%%%%%%%%%%%%%%%%%%%%%%%%%%%%%%

\section{Introduction}

Gauge mediated SUSY breaking (GMSB) models are one of the best motivated extensions of the Standard Model (SM) at the TeV scale. For instance, dangerous flavor-changing neutral currents are naturally suppressed, and the big-bang nucleosynthesis can work successfully without suffering from photo- and hadrodissociation by the gravitino decay. 

Recently, the ATLAS and CMS collaborations have reported excesses of events in the SM Higgs boson searches at the mass around $124-126\GeV$ \cite{HiggsDec13}. In the minimal SUSY Standard Model (MSSM), such a relatively heavy Higgs boson mass requires either a large trilinear coupling of the top squark or very large squark masses.
In GMSB models, since the trilinear coupling of the top squark is predicted to be small, a large SUSY-breaking mass scale is required to explain such a heavy Higgs boson. 

On the other hand, the experimental result of the muon anomalous magnetic moment (muon $g-2$) \cite{Bennett:2006fi} is deviated from the SM prediction at more than the $3\sigma$ level \cite{g-2_hagiwara}. The deviation may be explained by SUSY contributions, which are required to satisfy
\begin{equation}
  a_\mu({\rm SUSY}) = (26.1 \pm 8.0) \times 10^{-10}
\end{equation}
at the 1$\sigma$ level. 
However, if the soft mass scale is large, SUSY contributions to the muon $g-2$ become suppressed.
As a result, it is very difficult to explain both the Higgs boson mass of $124-126\GeV$ and the muon $g-2$ anomaly in GMSB models.

Recently, extensions of the GMSB models were studied in light of the Higgs boson mass and the muon $g-2$ \cite{arXiv:1108.3071,Endo:2011xq,Endo:2011gy,Evans:2012hg}. Some of them predict a large Higgsino mass term ($\mu$ term). Interestingly, large $\mu$ helps to enhance the SUSY contributions to the muon $g-2$, since the smuon--Bino diagram is proportional to the $\mu$ parameter \cite{Moroi:1995yh}. However, such a large $\mu$ parameter can destabilize the vacuum of the universe, since the trilinear coupling of the stau--Higgs term is enhanced (see \cite{Ratz:2008qh,Endo:2010ya,Hisano:2010re} for recent studies). In this work, the vacuum stability bound is studied in the following two GMSB extensions: i) the model with a large trilinear coupling of the top squark~\cite{Evans:2011bea,Evans:2012hg}, and ii) that with extra vector-like matters \cite{arXiv:1108.3071,arXiv:1108.3437,Endo:2011xq}. In both models, the trilinear coupling of the stau becomes large in the region where the Higgs is heavy. It will be found that the stability condition provides upper bounds on the gluino mass if combined with the muon $g-2$. We will also show that the whole parameter region which is favored by the Higgs mass and the muon $g-2$ is expected to be covered by the LHC experiment at $\sqrt{s} = 14\TeV$. The vacuum stability bound on the mSUGRA models with the vector-like matters~\cite{arXiv:1108.3071,Moroi:2011aa} will also be discussed, paying attention to the dark matter abundance.

%%%%%%%%%%%%%%%%%%%%%%%%%%%%%%
%%%%%%%%%%%%%%%%%%%%%%%%%%%%%%
%%%%%%%%%%%%%%%%%%%%%%%%%%%%%%
\section{Vacuum Stability Bound}
\label{sec:stability}

The stability of the ``ordinary'' vacuum, i.e., the electroweak-breaking vacuum in our universe, may be spoiled, if the trilinear coupling of the stau, 
\begin{eqnarray}
 {\cal L} \simeq \frac{g m_\tau}{2 M_W} \mu\tan\beta \stau_L^*\stau_R h^0 + {\rm h.c.},
 \label{eq:stau-trilinear}
\end{eqnarray}
is too large. Here we omit the sub-leading contributions which are not enhanced by the Higgsino mass parameter, $\mu$, nor a ratio of the Higgs VEVs, $\tan\beta$. As the trilinear coupling increases, an electric charge-breaking minimum becomes deeper. The stau trilinear coupling is therefore bounded from above so that the lifetime of the ordinary vacuum is longer than the age of the universe. 

The transition rate of the metastable vacuum is estimated by a semiclassical method, searching for so-called bounce solutions~\cite{Coleman:1977py}. In Ref.~\cite{Hisano:2010re}, an approximate formula for the bound on $\mu\tan\beta$  is  obtained by using multi-dimensional bounce configurations, including top--stop radiative corrections to the Higgs potential:
\begin{equation}
  \mu\tan\beta \lesssim 76.9 \sqrt{m_{\tilde\tau_L}m_{\tilde\tau_R}} + 38.7 (m_{\tilde\tau_L} + m_{\tilde\tau_R}) - 1.04 \times 10^4\,{\rm GeV},
  \label{eq:mubound}
\end{equation}
where $m_{\tilde\tau_L}$ and $m_{\tilde\tau_R}$ are soft scalar mass parameters for the left- and right-handed staus, respectively.
Although the correction to the Higgs potential depends on the masses of the stops, the bound (\ref{eq:mubound}) is affected only at the percent level~\cite{Hisano:2010re}~\footnote{The analysis in Ref.~\cite{Hisano:2010re} assumed the $A$-term in the trilinear coupling of the top squark, $A_t$, vanishing. Although a large $A_t$ can contribute through the top--stop loops, the following conclusion is considered to be less sensitive to it, because the corrections from the top--stop loops are small in Ref.~\cite{Hisano:2010re}.
}.

The above result is obtained in the limit of the zero temperature. The vacuum can transit through thermal effects in the early universe. The thermal decay rate of the false vacuum is usually estimated by following the method in Ref.~\cite{Linde:1981zj}. Evaluating the Higgs potential at the one-loop level, which includes the thermal potential coming from the top quark and electroweak gauge bosons, the stability bound on (\ref{eq:stau-trilinear}) can become more severe than the zero-temperature result (\ref{eq:mubound}) for a small stau mass region~\cite{Endo:2010ya}. However, the bound on the stau trilinear coupling (\ref{eq:stau-trilinear}) would be more severe only up to $\sim 10$\%~\cite{Endo:2010ya}, and hence we will adopt the bound (\ref{eq:mubound}) in the following models, which provides a conservative constraint on the parameter space.

The ordinary vacuum may be required to be absolutely stable if the vacuum expectation values of the scalar fields stayed away from the ordinary one in the early universe. In the following analysis, we will also show the constraint on the parameter space which is obtained by requiring the absolute stability of the vacuum (bearing in mind that the bound depends on the cosmological history).
The upper bound on (\ref{eq:stau-trilinear}) becomes more severe by about 50\% compared to the metastable one (\ref{eq:mubound})~\cite{Hisano:2010re}.

Finally, let us touch on another possibility of the vacuum instability. In a class of SUSY models, the trilinear coupling of the top squark is predicted to be large in order to raise the Higgs mass. Such a large trilinear coupling may spoil the vacuum stability in the stop--Higgs plane similarly to the stau--Higgs plane discussed above. The metastability bound has been studied in Ref.~\cite{Kusenko:1996jn} and obtained as $A_t^2 + 3\mu^2 < 7.5 (m_{\tilde t_L}^2 + m_{\tilde t_R}^2)$, where $A_t$ is a SUSY-braking trilinear coupling of the top squark defined as $V = -y_t A_t H_u^0 \tilde t_L^* \tilde t_R + {\rm h.c.}$. On the other hand, depending on the thermal history of the universe, the vacuum may be required to be absolute stable. The condition is found in Ref.~\cite{LeMouel:2001sf}, which we will use in the following analysis.

%%%%%%%%%%%%%%%%%%%%%%%%%%%%%%
%%%%%%%%%%%%%%%%%%%%%%%%%%%%%%
%%%%%%%%%%%%%%%%%%%%%%%%%%%%%%
\section{GMSB with Large $A_t$}
\label{sec:GMSBwithA}

In the simplest GMSB models, since the trilinear coupling of the top squark is suppressed at the messenger scale, it is difficult to realize the Higgs mass of $\gtrsim 124$\,GeV in the region where the muon $g-2$ discrepancy is explained. This difficulty can be ameliorated if the Higgs doublets mix with the messenger doublets~\cite{Evans:2011bea,Evans:2012hg}. This mixing induces a large trilinear coupling of the top squark, and hence the Higgs mass is enhanced. In this section, we study the vacuum stability bound in this model.

The model has six input parameters, $(\Lambda_{\rm mess}, M_{\rm mess}, N_{\rm mess}, \tan\beta, y'_t, \sgn(\mu))$, where $\Lambda_{\rm mess} \equiv F/M_{\rm mess}$ determines the soft mass scale.
The sign of $\mu$ is fixed to be positive in the following analysis, since otherwise the muon $g-2$ anomaly cannot be explained.
The Yukawa coupling $y'_t$ parametrizes the new superpotential,
\begin{equation}
  W = y'_t \Phi_{\bar L} Q_{L3} \bar T_R, 
\end{equation}
where $\Phi_{\bar L}$ is the leptonic component of the messenger field, and $Q_{L3}$  and $\bar T_R$ are the left- and right-handed top quark chiral multiplets, respectively.
This coupling induces corrections for the soft parameters.
The detailed mass spectrum is obtained in Refs.~\cite{Evans:2011bea,Evans:2012hg}.
Remarkably, the soft SUSY breaking parameters of the third generation squarks, including $A_t$ and $m_{\tilde t_{L,R}}^2$, receive extra threshold corrections.
Furthermore, the soft mass of the up-type Higgs gets a negative contribution, which essentially enhances $\mu$.

The Higgs mass and the muon $g-2$ in this model have been studied in Ref.~\cite{Evans:2012hg}.
It was found that the two conditions, $m_h > 124\GeV$ and $a_\mu ({\rm SUSY}) \gtrsim 10^{-9}$, are simultaneously satisfied when the messenger scale is low, i.e., $x \sim 0.4-0.5$ with $x \equiv \Lambda_{\rm mess}/M_{\rm mess}$, and the extra Yukawa coupling is large such as $y'_t \gsim 1$ at the messenger scale.
It was also argued that large $\tan\beta$ is favored to explain the muon $g-2$ anomaly.
Since $\mu$ becomes quite large, the vacuum stability bound in this model is of great importance as discussed in the previous section.

In Fig.~\ref{fig:gmsbA}, contours of the Higgs mass and the muon $g-2$ are displayed as a function of the physical gluino mass and $\tan\beta$.
The other parameters are fixed to be $(N_{\rm mess}, x, y'_t) = (1, 0.4, 1.3)$, $(2, 0.45, 1.2)$ and $(3, 0.5, 1.4)$, respectively in each plot. For $N_{\rm mess} = 2$ and 3, it is assumed that a pair of the messenger fields has a Yukawa interaction with the top.
The formulae of the soft parameters can be found in Refs.~\cite{Evans:2011bea,Evans:2012hg}.
In the numerical analysis, the mass spectrum is obtained by modifying {\tt SoftSUSY}~\cite{Allanach:2001kg}, and the Higgs mass and the muon $g-2$ are evaluated by {\tt FeynHiggs}~\cite{Hahn:2010te}. 
In the plots, the green region denotes $124\GeV < m_h < 126\GeV$, and the muon $g-2$ is explained in the orange (yellow) region within the $1(2)\sigma$ level.
It is found that the green band covers a wide parameter region because large $y'_t$ contributes to $A_t$. On the other hand, the soft mass scale, i.e., the gluino mass, is constrained by the muon $g-2$, and the region depends on $\tan\beta$.

The metastability bound is shown by the black solid lines in Fig.~\ref{fig:gmsbA}. This provides a severe upper bound on $\tan\beta$. 
Combined with the muon $g-2$, the gluino mass is bounded from above by the stability condition. This result is rather stable against variations of the model parameters, $x$ and $y'_t$. Around the parameter region where the Higgs mass is about $124\GeV$, the SUSY contributions to the muon $g-2$ do not depend much on $x$ and $y'_t$, though the Higgs mass is sensitive to them (see Ref.~\cite{Evans:2012hg} for the dependence on $x$ and $y'_t$). As a result, the gluino mass is found to be bounded as $m_{\tilde g} \lsim 1.3$ (0.9), 1.8 (1.4) and 2.1 (1.5)\,TeV for $N_{\rm mess} = 1$, 2 and 3, respectively, in the region where the muon $g-2$ is explained at  $2\sigma$ ($1\sigma$) level.

The absolute stable bound is also displayed by the black dashed line as labelled by $\tilde\tau$ in Fig.~\ref{fig:gmsbA}. If the thermal history of the universe requires the absolute stability of the ordinary vacuum, the constraint on $\tan\beta$ is found to be very stringent. As a result, the gluino mass becomes tightly limited such as as $m_{\tilde g} \lsim 1.1$ (0.8), 1.5 (1.2) and 1.7 (1.3)\,TeV for $N_{\rm mess} = 1$, 2 and 3, respectively, for the muon $g-2$ explanation at $2\sigma$ ($1\sigma$). The stability constraint gives severe bounds on the superparticle masses, which is essential for the collider searches for the superparticles. We will discuss LHC searches later.

Let us mention some other constraints. 
In the figures, the trilinear coupling of the top squark is large due to sizable $y'_t$, which may destabilize the vacuum in the stop--Higgs plane. We have checked that the metastability condition is satisfied in the whole parameter region in Fig.~\ref{fig:gmsbA}. 
On the other hand, the absolute stability bound is shown by the black dashed line with a label, $\tilde t$, for $N_{\rm mess} = 2$ in Fig.~\ref{fig:gmsbA}. Above the line, there is a deeper vacuum than the ordinary one. Thus, most of the parameter region is potentially excluded by the constraint in this case, though whether it applies or not depends on the thermal history.

A large trilinear coupling of the top squark also enhances a SUSY contribution to the branching ratio of $b \to s \gamma$. The experimental result, ${\rm Br}(\bar B \to X_s \gamma)^{\rm exp} = (3.55 \pm 0.24 \pm 0.09) \times 10^{-4}$~\cite{Asner:2010qj}, agrees with the SM prediction, ${\rm Br}(\bar B \to X_s \gamma)^{\rm SM} = (3.15 \pm 0.23) \times 10^{-4}$~\cite{Misiak:2006zs}, very well, where $\bar B$ represents $\bar B^0$ or $B^-$. Hence, the SUSY contributions are limited. We evaluated the SUSY contributions of the model at the NLO level by {\tt SusyBSG}~\cite{Degrassi:2007kj}. In addition to the experimental and SM uncertainties, errors of $10\%$ are taken into account both for the SUSY and charged Higgs contributions, respectively (see e.g.~\cite{Degrassi:2007kj}). In the minimal flavor violation, the dominant contribution of SUSY comes from the chargino--stop loop diagram and is enhanced by large $\tan\beta$ as well as $A_t$. Requiring that the SUSY contribution is within the $2\sigma$ range of the SM and experimental results, $b \to s \gamma$ provides a constraint, which is displayed by the gray shaded region in the panel of $N_{\rm mess} = 2$ and that with small $\tan\beta$ for $N_{\rm mess} = 3$ in Fig.~\ref{fig:gmsbA}. 

\subsection{LHC phenomenology}

The LHC constraints and prospects of the model have been discussed in Ref.~\cite{Evans:2012hg}. Here, we discuss the LHC phenomenology in light of the vacuum stability bound obtained above. We will also extend the analysis including the cases of prompt NLSP decays.

\begin{figure}[t]
\vspace{-20pt}
\begin{flushleft}
\includegraphics[scale=0.65]{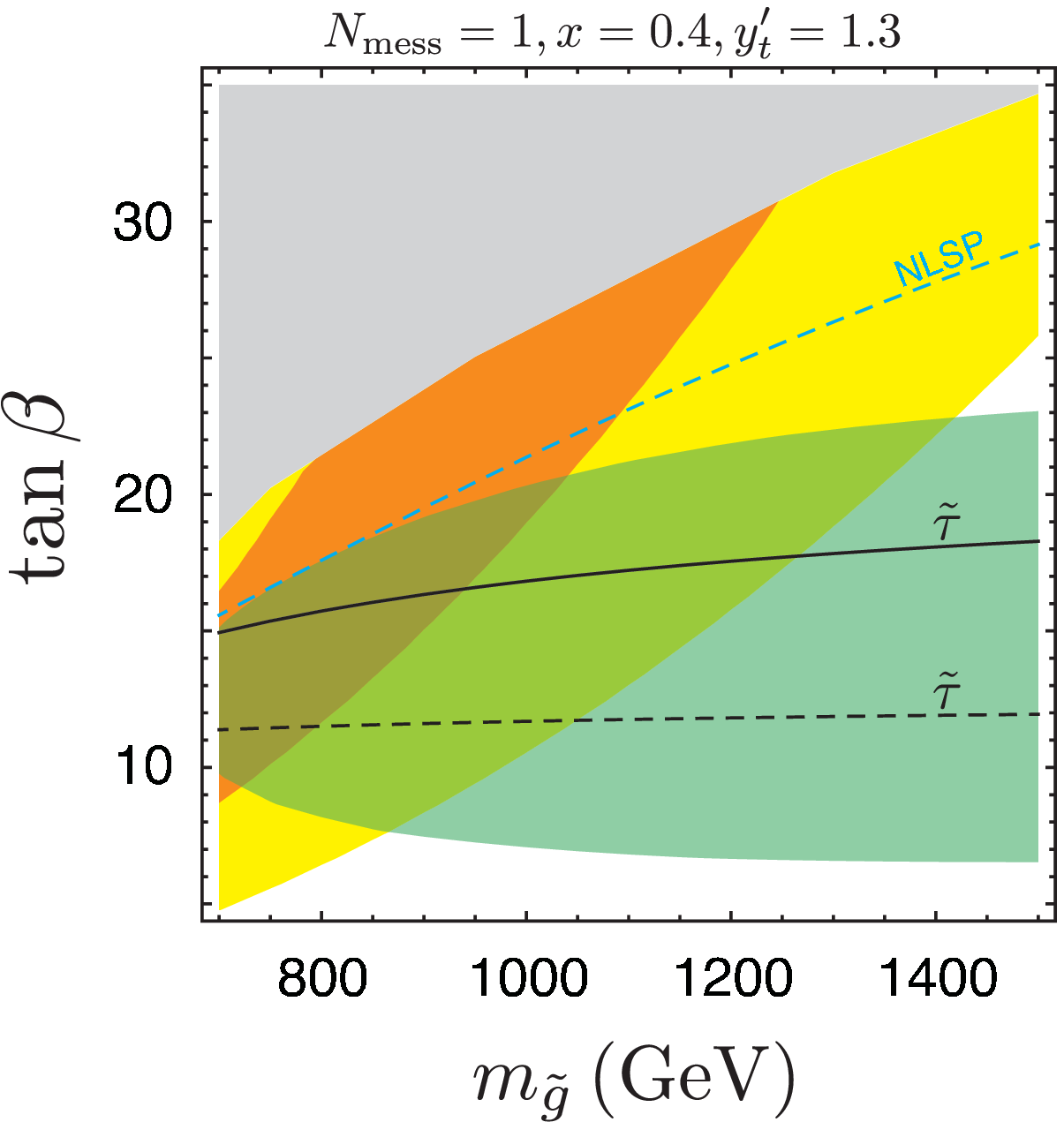}\hspace*{5mm}
\includegraphics[scale=0.65]{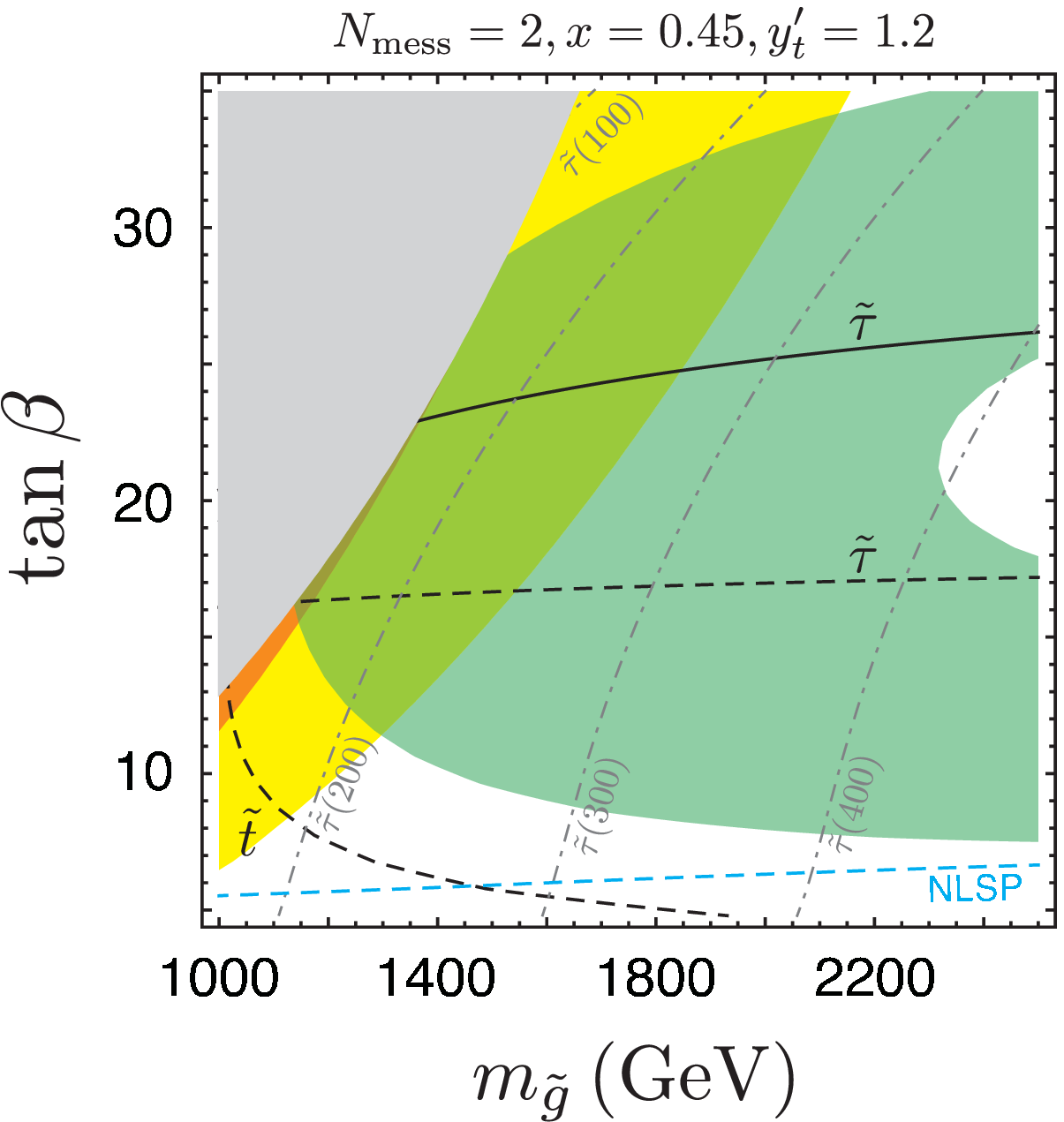}\\\vspace*{5mm}
\includegraphics[scale=0.65]{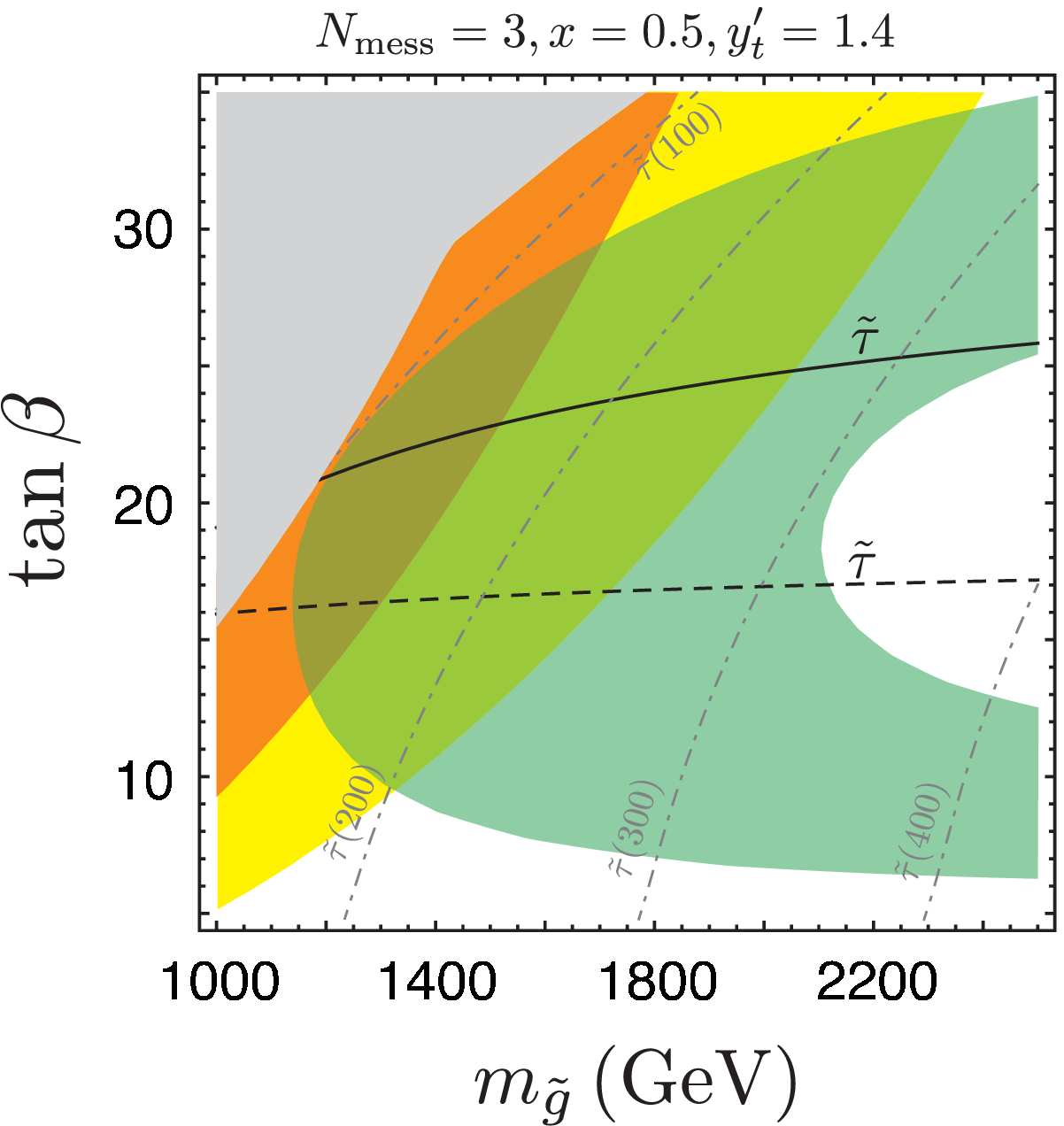}
\caption{
Contours of the Higgs boson mass of $124-126\GeV$ (green region) and the muon $g-2$ at the $1(2)\sigma$ level (orange(yellow) region) in the GMSB model with a large trilinear coupling of the top squark. The sign of $\mu$ is set to be positive. The black solid and dashed lines labelled by $\tilde \tau$ represent the upper bounds on $\tan\beta$ obtained from the meta- and absolute stabilities of the vacuum along the stau-Higgs direction, respectively.
The region above the black solid lines are excluded.
For $N_{\rm mess}=2$, 
the black dashed line labelled by $\tilde t$ corresponds to the upper bound on $\tan\beta$ obtained from the absolute stability of the vacuum along the stop-Higgs direction. 
For $N_{\rm mess}=1$ and 2, above the light blue dashed line, the stau is the NLSP, while the neutralino is lighter below it. 
For $N_{\rm mess}=2$ and 3, the stau mass is displayed by gray dot-dashed lines. The gray shaded regions are excluded by the tachyonic stau and $b \to s\gamma$. }
\label{fig:gmsbA}
\end{flushleft}
\end{figure}

The collider signature of the SUSY events is determined by the next-to-lightest superparticle (NLSP), since the lightest superparticle (LSP) is the gravitino. Above the light blue dashed line the stau is the NLSP in the panels of $N_{\rm mess} = 1$ and 2 in Fig.~\ref{fig:gmsbA}, while the neutralino becomes lighter below it. In the case of $N_{\rm mess} = 3$, the stau is the NLSP in the whole parameter region. The gray dot-dashed contours denote the stau mass. As $\tan\beta$ increases, the lightest stau mass decreases. The gray shaded region in the $N_{\rm mess} = 1$ panel and that in a large $\tan\beta$ region for $N_{\rm mess} = 3$ are excluded by a tachyonic stau. 

The collider signature depends on the lifetime of the NLSP.
Its decay length depends on the mass of the gravitino and NLSP as 
$c\tau \sim 10\mu {\rm m} (m_{3/2}/1\,{\rm eV})^2 ( m_{\rm NLSP} / 100\GeV )^{-5}$.
The gravitino mass is determined by the magnitude of the total SUSY breaking scale, $F_{\rm total}$, which is equal to or larger than the $F$ of the messenger sector, $F=\Lambda_{\rm mess} M_{\rm mess}$. 
As discussed in Ref.~\cite{Evans:2012hg}, the vacuum stability for the messenger sector tends to require $F\ll F_{\rm total}$, and thus a long-lived NLSP is favored. For completeness, we also discuss the case of $F\simeq F_{\rm total}$, where the NLSP can decay promptly~\footnote{If the gravitino mass is mildly light, the NLSP can decay inside the detectors. Then, SUSY events include signatures with kinks or non-pointing photons. Such cases need to be studied elsewhere. }.
In the whole parameter region of Fig.~\ref{fig:gmsbA}, $F/(\sqrt{3}M_P) \sim (1-10)$\,eV, where $M_P$ is the reduced Plank mass scale. The NLSP either decays promptly or is long-lived, depending on the gravitino mass.
In the following we discuss SUSY searches in the LHC for both cases.
In the following, the Monte Carlo simulation relies on {\tt Pythia}~\cite{Pythia6.4} to obtain the event shape, and the detector simulation is based on {\tt PGS}~\cite{PGS4}. The cross sections are evaluated by {\tt Prospino}~\cite{Beenakker:1996ed} at the NLO and LO levels for the colored superparticle productions and the electroweak processes, respectively.

When the messenger number is $N_{\rm mess} = 1$, the Bino is the NLSP in the whole parameter region where the metastability bound is satisfied, i.e., below the black solid line in Fig.~\ref{fig:gmsbA}. Since the Bino is a neutral particle, the SUSY events are associated by a large missing transverse energy, $E_T^{\rm miss}$, if the Bino is long-lived. Thus, the hadronic channels with large missing energy are suitable for the detection, and the relevant production channels are those of the gluino and squarks, $pp \to \tilde g\tilde g$, $\tilde g\tilde q$ and $\tilde q\tilde q$. Applying the high mass cut used in the ATLAS search~\cite{Aad:2011ib}, it is found that the gluino mass is required to be larger than around $900\GeV$ in Fig.~\ref{fig:gmsbA}. Thus a corner of the parameter space is left to explain the muon $g-2$ anomaly at the 1 $\sigma$ level.

It is commented that this search is similar to those for the CMSSM models~\cite{Aad:2011ib,CMS:SUS-11-004-pas,CMS:SUS-11-003-pas,CMS:SUS-11-008-pas}. The difference is that the sleptons are likely to be much lighter in the GMSB models. If the slepton mass is close to that of the NLSP, the leptons which are generated in decay chains of the colored superparticles are likely to be soft. Thus,  the number of events which pass through the cut condition, the lepton veto, increases compared to that of  CMSSM.

If the Bino decays promptly, photons are likely to be generated when the Bino decays into the gravitino. Since the gravitino provides a missing energy, the di-photon with large $E_T^{\rm miss}$ is a promising channel for the selection. Such a signature has been studied in ATLAS~\cite{Aad:2011zj} and CMS~\cite{CMS:SUS-11-009-pas}. The main production channels are electroweak productions of the charginos, neutralinos and sleptons. Applying the same cut as the ATLAS analysis~\cite{Aad:2011zj}, the bound is obtained as $m_{\tilde g} \gsim 1.1\TeV$ for $N_{\rm mess} = 1$ in Fig.~\ref{fig:gmsbA}. Thus, if the Bino is the NLSP with a short lifetime, the LHC has already excluded the region which is allowed by the vacuum stability and is consistent with the muon $g-2$ at the 1$\sigma$ level.

As the messenger number increases, the stau tends to be lighter than the Bino. For $N_{\rm mess} = 2$ and 3, the stau becomes the NLSP in most of the relevant parameter region as shown in Fig.~\ref{fig:gmsbA}. If the gravitino is much heavier than $O(1)$\,eV, the stau is long-lived and leaves a charged track in the detectors. Heavy stable charged particles are searched for in ATLAS~\cite{Aad:2011hz} and CMS~\cite{CMS:EXO-11-022-pas}. The CMS analysis was updated very recently by using the integrated luminosity, 4.7\,fb$^{-1}$~\cite{CMS:update2012}. If the updated analysis is done as the previous one~\cite{CMS:EXO-11-022-pas}, the constraint is found to be $m_{\tilde g} \gsim 1.9\TeV$ both for $N_{\rm mess} = 2$ and 3 in Fig.~\ref{fig:gmsbA}, where the main production channels are electroweak productions of the charginos, neutralinos and sleptons. The bound on the stau mass by the stau direct production is weaker than the ino productions. As a result, most of the parameter space which is favored by the muon $g-2$ and the vacuum stability has been already excluded by the long-lived stau searches. 

If the stau decays promptly, the most relevant channel is associated by the di-lepton with large $E_T^{\rm miss}$. The bound has been studied in the framework of the general gauge mediation, and the same-sign di-lepton channel provides the severest bound~\cite{Kats:2011qh}. The ATLAS analyzed the opposite-sign di-lepton channel in the GMSB models~\cite{Aad:2011cwa}. In both cases, the gluino mass $> 1\TeV$ is large enough to avoid the detections. Thus, no constraints are obtained for $N_{\rm mess} = 2$ and 3 in Fig.~\ref{fig:gmsbA} as long as the stau is the NLSP and decays promptly.

Finally, let us mention future prospects for the discovery. When the NLSP Bino is long-lived, the signal is very similar to the CMSSM. Since the gluino and squark masses determine the production cross section, the discovery potential can be read off by translating these masses into $(m_0, m_{1/2})$. If the LHC will run at $\sqrt{s} = 7\TeV$ during 2012, the gluino mass region of $\gsim 1.2\TeV$ is expected to be accessed for the integrated luminosity of 30\,fb$^{-1}$~\cite{Baer:2011aa}. The sensitivity would be better if the energy is upgraded to be $\sqrt{s} = 8\TeV$. All the region favored by the muon $g-2$ is possibly covered by the LHC at $\sqrt{s} = 14\TeV$ with the integrated luminosity of $O(1-10)\,{\rm fb}^{-1}$~\cite{Zhukov:2006ay,Baer:2009dn}. On the other hand, if the stau is the NLSP and decays promptly, the multi-lepton channels are considered to be suitable discovery channels. Based on the study of the discovery potential of the low-scale GMSB, the gluino with a mass of around $2\TeV$ is expected to be discovered for $O(1-10)\,{\rm fb}^{-1}$ at $\sqrt{s} = 14\TeV$~\cite{Nakamura:2010faa}. Thus, the whole parameter region which is favored by the muon $g-2$ and constrained by the metastability bound can be covered by the future LHC experiments. 

%%%%%%%%%%%%%%%
%%%%%%%%%%%%%%%

%%%%%%%%%%%%%%%%%%%%%%%%%%%%%%
%%%%%%%%%%%%%%%%%%%%%%%%%%%%%%
%%%%%%%%%%%%%%%%%%%%%%%%%%%%%%
\section{The Model with Vector-like Matters}

We consider an extended model of the MSSM with vector-like matters. The extra matters are introduced as complete SU(5) multiplets, ${\bf 10}=(Q', U', E')$ and ${\bf\overline{10}}=(\bar{Q}', \bar{U}', \bar{E}')$. The details are found, e.g.~in \cite{arXiv:1108.3071}. The model is highly motivated since it can explain the muon $g-2$ and the Higgs mass, $m_h \simeq 125$ GeV, simultaneously both for the GMSB~\cite{arXiv:1108.3071,Endo:2011xq} and mSUGRA scenarios~\cite{arXiv:1108.3071,Moroi:2011aa}.
The superpotential is given by
\begin{eqnarray}
W = W_{\rm MSSM} + Y' Q' H_u U' + Y'' \bar{Q}' H_d \bar{U}' + M_{Q'} Q' \bar{Q}' + M_{U'} U' \bar{U}' + 
M_{E'} E' \bar{E}', \label{eq:super}
\end{eqnarray}
where $W_{\rm MSSM}$ is the superpotential for the MSSM matters. In the presence of the extra up-type (s)quarks, the Higgs mass is lifted by their radiative corrections. This becomes significant when the Yukawa coupling is as large as $Y' \simeq 1$, and the SUSY invariant masses, $M_{Q'}$ and $M_{U'}$, are in the TeV scale~\cite{Moroi:1991mg,Babu:2004xg+2008ge,Martin:2009bg}. A model was proposed in Ref.~\cite{Asano:2011zt}, which can naturally explain the existence of such TeV scale vector-like matters.
The extra Yukawa coupling, $Y'$, has an infrared fix point at $Y'({\rm EW}) \simeq 1$~\cite{Martin:2009bg}.

The vector-like matters raise the $\mu$ parameter.
When there are extra matters, the soft mass squared of the up-type Higgs, $m_{H_u}^2$, is driven down at the electroweak scale compared to that of the MSSM. This is due to the renormalization group (RG) evolution of $m_{H_u}^2$. The additional contribution from the extra squarks to the RG equation is given by
\begin{eqnarray}
(16\pi^2)\frac{d m_{H_u}^2}{d \ln Q} \ni 6 Y'^2 (m_{H_u}^2 + m_{Q'}^2 + m_{U'}^2 +A'^2 )
\end{eqnarray}
at one-loop level, where $Q$ is a renormalization scale, $m_{Q'}$ and $m_{U'}$ are soft masses of the extra squarks, and  $A'$ is a scalar trilinear coupling which corresponds to the second term of the right-hand side of Eq.~({\ref{eq:super}). Since $Y'\simeq 1$ is required to lift up the Higgs mass, this correction becomes sizable. From the electroweak symmetry breaking (EWSB) condition, $\mu$ is found to be large, which can induce a deep electric charge-breaking minimum depending on $\tan\beta$ and the stau masses as discussed in Sec.~\ref{sec:stability}.

It is commented that the trilinear coupling of the top squark is small in the extra matter models. This is because large $Y'$ suppresses $A_t$ during the RG evolution. Thus, $A_t$ at the weak scale is insensitive to the input at a high scale and has an infrared fixed point at a small value~\cite{Martin:2009bg, Endo:2011xq}. This feature is distinct from the model in Sec.~\ref{sec:GMSBwithA}, where large $A_t$ potentially spoils the absolute stability condition and can suffer from the $b \to s \gamma$ bound. The models with the vector-like matters are free from these constraints.

In the following two subsections, the Higgs mass and the muon $g-2$ are evaluated under the stability condition both in the GMSB and mSUGRA scenarios. The mass spectrum of the SUSY particles are evaluated by {\tt SuSpect}~\cite{Djouadi:2002ze}, which is modified to include the contributions to renormalization group equations from the vector-like matters at the two-loop level. The MSSM contributions to the Higgs mass is calculated by {\tt FeynHiggs} at the NLO level, and those from the vector-like matters are at the one-loop level. The muon $g-2$ is also calculated by {\tt FeynHiggs}. In the mSUGRA, we calculate the relic abundance of the lightest neutralino by {\tt micrOMEGAs}~\cite{Belanger:2006is}. The collider Monte Carlo is simulated by the same packages as those in Sec.~\ref{sec:GMSBwithA}, {\tt Pythia}, {\tt PGS} and {\tt Prospino}. In the following analysis, we assume $Y''\simeq 0$, since otherwise the extra down-type squark reduces the SUSY contributions to the Higgs mass (see Ref.~\cite{Endo:2011xq}). 

\subsection{GMSB models with vector-like matters}

First we consider the simplest GMSB model, which is parametrized by $\Lambda_{\rm mess}$, $M_{\rm mess}$ and $N_{\rm mess}$ as well as $\tan\beta$ and $\sgn(\mu)$. Here we set the messenger number as $N_{\rm mess} = 1$, since larger $N_{\rm mess}$ spoils the perturbative gauge coupling unification, unless the messenger scale is high enough. 

In Fig.~\ref{fig:gmsb}, the Higgs mass, the muon $g-2$ and the stability bounds are shown in the $m_{\tilde{g}}$--$\tan\beta$ plane for various messenger scales. The green band shows the Higgs mass of $124\, {\rm GeV} < m_h^0 < 126\, {\rm GeV}$ for $M_{Q'}=M_{U'}=0.6$ and $1.0$ TeV. In the orange (yellow) region, the muon $g-2$ discrepancy is explained at the 1$\sigma$ (2$\sigma$) level. 

The vacuum stability gives stringent constraints on the soft mass scale. Above the black solid line, the region is excluded by the metastability condition (\ref{eq:mubound}). If the bound is combined with the muon $g-2$, the soft mass scale, which is represented by the gluino mass in Fig.~\ref{fig:gmsb}, is constrained from above. From (\ref{eq:mubound}), the upper bound is obtained as $m_{\tilde{g}}< 1.2\ (1.7)$\,TeV for the muon $g-2$ at the 1$\sigma\, (2 \sigma)$ level when the messenger scale is $10^6$\,GeV. Also, in the region above the black dot-dashed line, the ordinary vacuum is not absolutely stable but becomes metastable. This gives stronger upper bound on $\tan\beta$. The bound is found to be $m_{\tilde{g}}< 1.0\ (1.3)$\,TeV for the muon $g-2$ at the 1$\sigma\, (2 \sigma)$ level and $M_{\rm mess} = 10^6\GeV$. For both constraints, the upper bound on the gluino mass becomes severer if $M_{\rm mess}$ is larger,

In Fig.~\ref{fig:gmsb}, the NLSP is the lightest stau (neutralino) above (below) the blue dashed line. In all the panels of Fig.~\ref{fig:gmsb}, the NLSP is likely to be long-lived, since $M_{\rm mess}$ is larger than $10^6\GeV$. When the stau is the NLSP, the parameter regions below the black solid line and above the blue dashed one seem to survive the metastability condition especially for low $M_{\rm mess}$. However, those regions have been already excluded by the searches for the heavy stable charged particle in the LHC as studied in Ref.~\cite{Endo:2011xq}. On the other hand, if the messenger scale is as low as $\sim 10^5\GeV$, the gravitino mass could be of $O(1-10)$\,eV, and thus, the stau decays promptly. Then direct search bounds from the LHC are weak, and thus, the region with $m_{\tilde{g}} \gsim 1$\,TeV is still available (see \cite{Kats:2011qh,Aad:2011cwa}). Since the gluino mass is tightly bounded by the metastability condition, the whole regions are expected to be covered by searches for the multi-lepton events in early stage of the LHC at $\sqrt{s} = 14\TeV$~\cite{Nakamura:2010faa}. The regions below the black dot-dashed line already excluded the stau NLSP.

When the NLSP is the neutralino, long-lived neutralinos contribute to the missing transverse energy at the colliders. If the high mass cut of the ATLAS search~\cite{Aad:2011ib} is applied, it is found that the gluino mass is required to be larger than around $750\GeV$ in all the panels of Fig.~\ref{fig:gmsb}. Thus, a large part of the parameter space is still valid against the collider searches. According to the discussion in \cite{Endo:2011xq}, the whole parameter region with $m_{\tilde{g}}< 1.7\TeV$ is expected to be covered in future at the LHC if the integrated luminosity reaches $O(10-100)\invfb$ at $\sqrt{s} = 14\TeV$.

If the gravitino mass is $O(1-10)$\,eV, the neutralino NLSP decays promptly in the detectors. The di-photon events with a large missing transverse energy are searched for and provide a constraint on the SUSY particle masses. If the ATLAS result~\cite{Aad:2011zj} is applied, the bound is obtained as $m_{\tilde g} \gsim 1.2\TeV$ for $M_{\rm mess} = 2 \times 10^5\GeV$, in which case the mass spectrum is almost same as those of $M_{\rm mess} = 10^6\GeV$ in Fig.~\ref{fig:gmsb}. Thus, most of the parameter region which is favored by the muon $g-2$ at the 1$\sigma$ level has been excluded, and the whole 2$\sigma$ region of the muon $g-2$ is expected to be accessible in the LHC soon at $\sqrt{s} = 14\TeV$~\cite{Nakamura:2010faa}.

%%%%%%%%%%%%%%%
\begin{figure}[t]
\begin{center}
\includegraphics[scale=1.05]{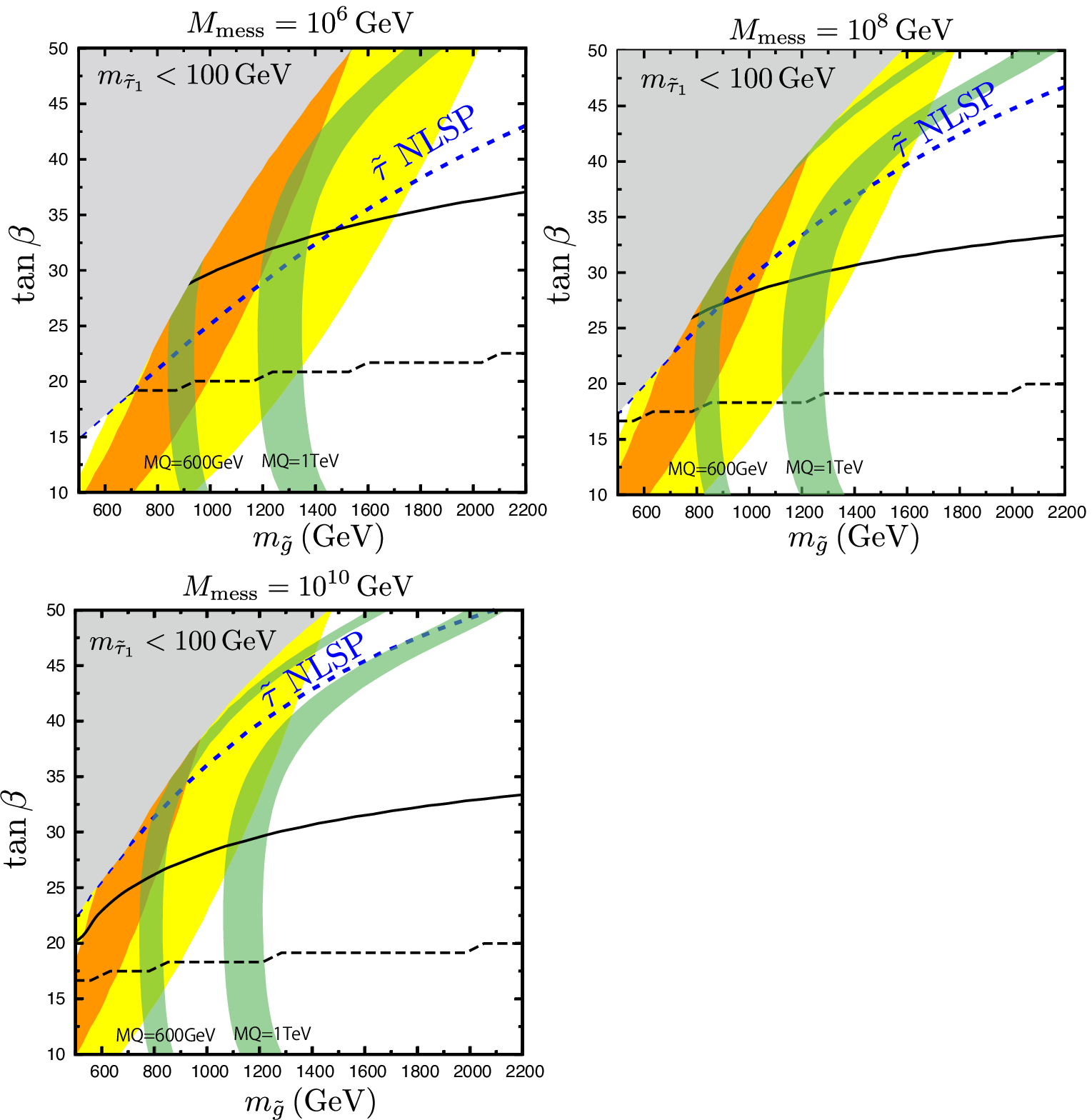}
\caption{Contours of the Higgs mass, the muon $g-2$ for various messenger scales in the GMSB model with extra vector-like matters. The region consistent with 124\,GeV $< m_h <$ 126\,GeV is shown as the green band for $M_{Q'}(=M_{U'})=600\GeV$ and 1\,TeV. The orange (yellow) region is consistent with the muon $g-2$ at the 1$\sigma$ (2$\sigma$) level. On the blue dashed line, masses of the lightest neutralino and stau are equal to each other. Above (bellow) the line, the lightest stau (neutralino) is the NLSP.
Above the black dot dashed line, the ordinary vacuum is metastable. 
The region above the black solid line is excluded by the vacuum instability. 
In the gray shaded region the stau mass becomes $m_{\tilde{\tau}_1} \lesssim 100$\,GeV.
  }
\label{fig:gmsb}
\end{center}
\end{figure}
%%%%%%%%%%%%%%%

%%%%%%%%%%%%%%%%%%%%%%%%%%%%%%%%%%%%%%%%%%%%%%%%%%%%%%%%%%%%%%%%%%%%
\subsection{mSUGRA models with vector-like matters}

Next, we consider the mSUGRA boundary condition, which is characterized by the following five parameters: the universal soft scalar mass $m_0$, the universal gaugino mass $m_{1/2}$, the universal trilinear coupling $A_0$, $\tan\beta$ and $\sgn(\mu)$ at the GUT scale. Since $A_t$ is suppressed at the weak scale and the other $A_i$'s are irrelevant for the Higgs mass and the muon $g-2$, we set $A_0=0$ in numerical analysis. We also take $\sgn(\mu) = +1$, which is favored by the muon $g-2$. 

The contours of the Higgs mass and the muon $g-2$ are shown in Fig.~\ref{fig:SUGRA1} and Fig.~\ref{fig:SUGRA2}, where the definitions of the regions and lines are same as Fig.~\ref{fig:gmsb}. In Fig.~\ref{fig:SUGRA1}, $\tan\beta$ are varied among the panels, while $m_0$ is set to be zero in Fig.~\ref{fig:SUGRA2}. The metastability bound from the stau is represented by the black solid line. This constraint becomes severer as $\tan\beta$ is larger. In addition, the region with larger $m_0$ avoids the constraint. The latter feature is understood as follows: the stau mass is approximately proportional to $m_0$, while the size of $\mu$ is less sensitive to it and rather determined by the gaugino (gluino) mass $m_{1/2}$ (see \cite{Martin:2009bg}) in most of the parameter region of our interest. Thus, the bound (\ref{eq:stau-trilinear}) is satisfied in a larger $m_0$ region. The absolute stability bound displayed by the black dot-dashed line is understood similarly. 

From Fig.~\ref{fig:SUGRA1}, it is found that when $\tan\beta$ is large and $m_0$ is small, the muon $g-2$ can be of order $10^{-9}$ easily in the region where the Higgs mass is $124-126\GeV$. However, such regions are likely to be excluded by the metastability condition.  Moreover, if the ordinary vacuum is required to be absolutely stable, the lower bound on $m_0$ as well as the upper bound on $\tan\beta$ becomes tighter. As a result, it is more difficult to satisfy both $m_h = 124-126\GeV$ and explain the muon $g-2$ anomaly at the 2$\sigma$ level unless $\tan\beta$ is properly small.

In Fig.~\ref{fig:SUGRA2}, we show the Higgs mass and the muon $g-2$ on the $m_{\tilde g}$--$\tan\beta$ plane. Here $m_0$ is set to be zero so that the SUSY contributions to the muon $g-2$ become as large as possible. Note also that the physical gluino mass is smaller by $O(10)$\% than $m_{1/2}$ in the extra vector-like matter models~\cite{Martin:2009bg}. It is found that the gluino mass is bounded from above if the stability bounds on $\tan\beta$ are combined with the muon $g-2$. The metastability bound is found to be $m_{\tilde{g}} \lesssim 0.8\ (1.1)$\,TeV when the muon $g-2$ deviation is explained at the 1$\sigma$ (2$\sigma$) level. If the absolute stability is imposed, the gluino is required to satisfy $m_{\tilde{g}} \lesssim 600\ (850)$\,GeV for the 1$\sigma$ (2$\sigma$) level of the muon $g-2$. 

In the mSUGRA models, the Bino is the LSP. The searches for the multi-jets events with a large missing transverse energy provide a constraint for the gluino and squark masses. In the current model, the squark mass is related to the gluino one, and the gluino mass is required to be larger than around $750\GeV$ if the high mass cut of the ATLAS search~\cite{Aad:2011ib} is applied. Thus, most of the region which is favored by the muon $g-2$ at the 1$\sigma$ level has been excluded already, and is expected to be covered by the LHC soon~\cite{Baer:2011aa}. Moreover, since the vector-like matter is required to be relatively light in order to raise the Higgs mass sufficiently, the extra matters are expected to be discovered by searches for the fourth generation quarks at the LHC~\cite{Endo:2011xq}. On the other hand, if the muon $g-2$ allows the 2$\sigma$ discrepancy, the upper bound on the gluino mass is relaxed. The gluino mass of $m_{\tilde{g}} = 1.1$\,TeV is in reach of the LHC experiment at $\sqrt{s} = 7-8\TeV$ with the integrated luminosity of $\sim 10\invfb$~\cite{Baer:2011aa}. Thus, the whole favored region with the muon $g-2$ at the 2$\sigma$ level will be checked in the near future at the LHC. 

Finally let us mention the relic abundance of the lightest neutralino. 
The blue solid lines denote a contour of the neutralino abundance with $\Omega_{\rm CDM}h^2 \simeq 0.1$ in Fig.~\ref{fig:SUGRA1} and Fig.~\ref{fig:SUGRA2}. It is noticed that the region exists for small $m_0$ if $\tan\beta$ is large, where the coannihilation works efficiently for the neutralino LSP. Such parameters are likely to spoil the stability of the vacuum as mentioned above. Consequently, the coannihilation region is found for $m_{\tilde{g}} \lesssim 600\GeV$ from Fig.~\ref{fig:SUGRA2}. However, the region has been already excluded by the LHC. Thus, in order for the neutralino LSP to explain the measured abundance of the dark matter, late-time entropy productions should occur after the neutralino decoupled from the thermal bath.

%%%%%%%%%%%%%%%
\begin{figure}[t]
\begin{center}
\includegraphics[scale=1.1]{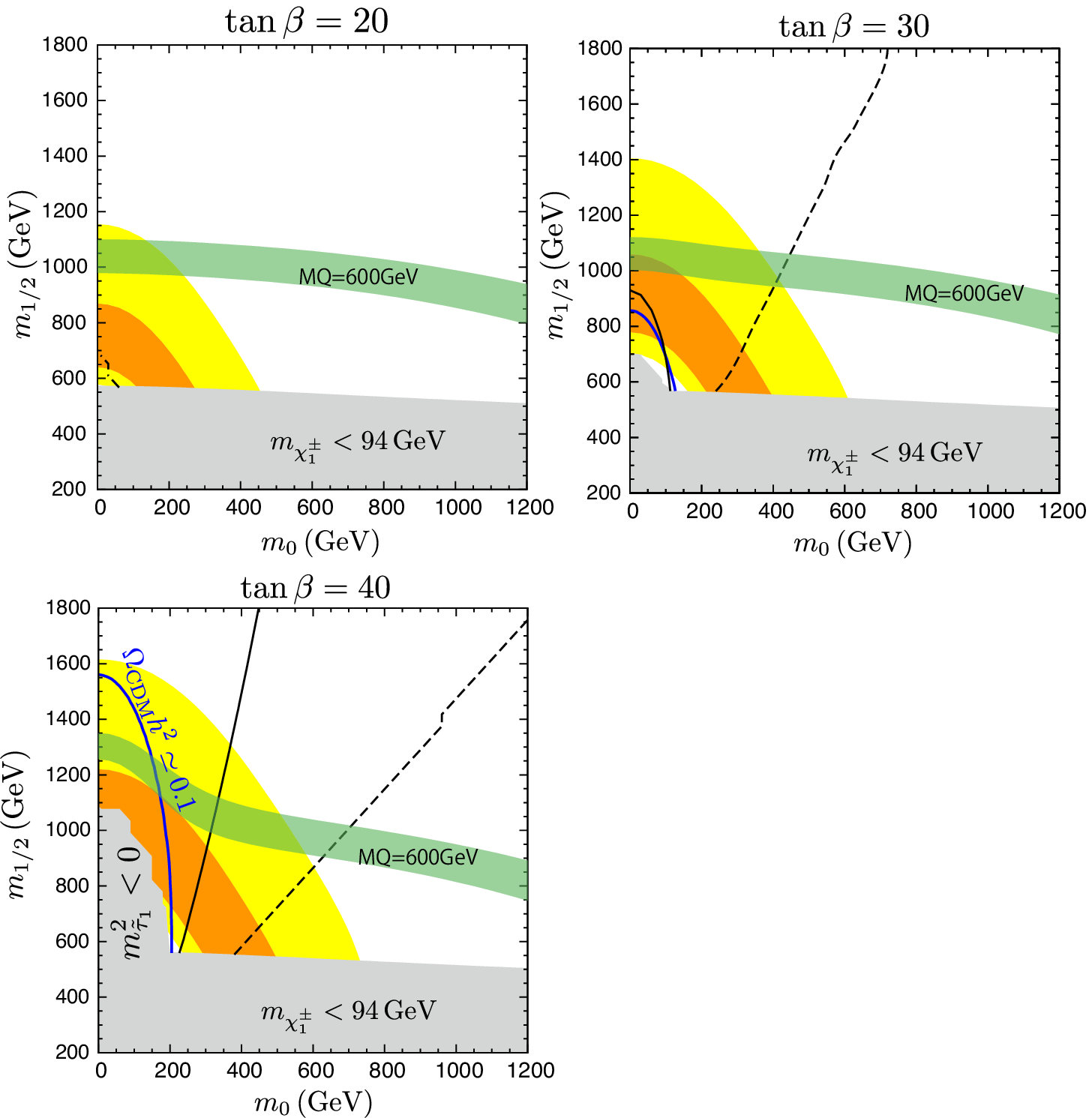}
\caption{Contours of the Higgs mass, the muon $g-2$ and the relic abundance of the dark matter in  mSUGRA models with extra vector-like matters. 
The region/lines are same as in Fig.~\ref{fig:gmsb} except for the blue solid line, which denotes the contour of the relic abundance of the lightest neutralino, $\Omega_{\rm CDM}h^2\simeq 0.1$. In the left region of the black dot dashed line, the ordinary vacuum is metastable. In the left region of the black solid line is excluded by the vacuum stability condition. The other mSUGRA parameters are taken as $A_0=0$ and $\sgn(\mu)$=1. The gray region is excluded, since the stau is tachyonic or the chargino is too light.
  }
\label{fig:SUGRA1}
\end{center}
\end{figure}
%%%%%%%%%%%%%%%

%%%%%%%%%%%%%%%
\begin{figure}[t]
\begin{center}
\includegraphics[scale=1.05]{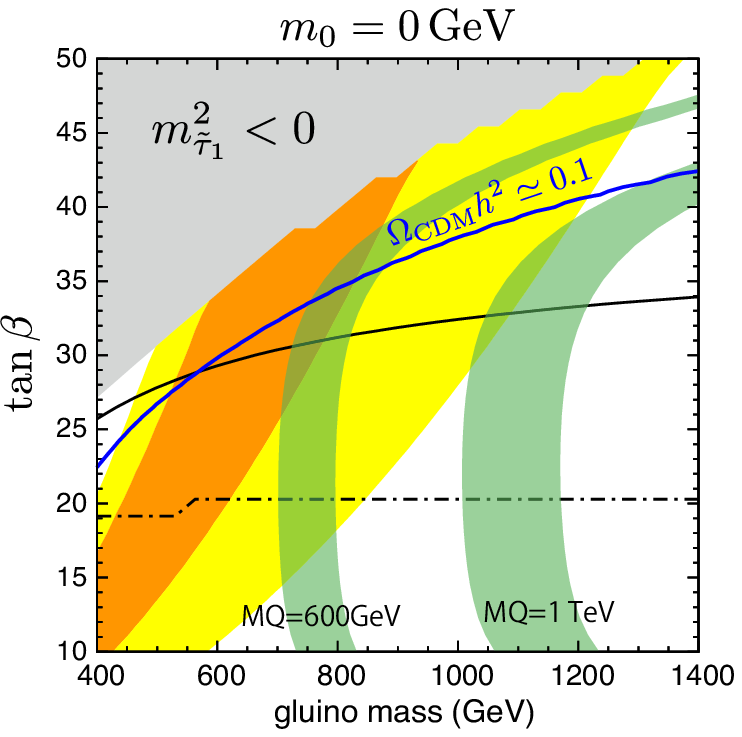}
\caption{The regions/lines are same as in Fig.~{\ref{fig:SUGRA1}}. Here, $m_{1/2}$ and $\tan\beta$ are varied with setting $m_0=0$, $A_0=0$ and ${\rm sign}(\mu)$=1. The region above the black solid line is excluded. The gray region is excluded, since the stau is tachyonic. 
  }
\label{fig:SUGRA2}
\end{center}
\end{figure}
%%%%%%%%%%%%%%%

%%%%%%%%%%%%%%%%%%%%%%%%%%%%%%
\section{Conclusion}

The simple GMSB models are disfavored by the muon $g-2$ in light of the recent results of the Higgs boson search. 
Extensions of the models were recently explored to accommodate the muon $g-2$ to the Higgs mass of $124-126\GeV$.
As a characteristic feature, some models predict a large $\mu$ term. In this work, the vacuum stability constraint from the stau was studied.
In particular, we studied two GMSB extensions: i) the model with a large trilinear coupling of the top squark, and ii) that with the vector-like matters. In both models, the vacuum stability condition was found to provide upper bounds on the gluino mass if the constraint is combined with the muon $g-2$. Based on the gluino mass upper bound, the discoveries at the LHC were also discussed. It was concluded that the whole parameter region obtained by the metastability bound and the muon $g-2$ is expected to be covered by LHC at $\sqrt{s} = 14\TeV$. 

This analysis was also applied to the mSUGRA models with the vector-like matters. 
%The region where the Higgs mass becomes $124-126\GeV$ and the muon $g-2$ anomaly is explained at the 2$\sigma$ level exists for low $\tan\beta$.
% 
The whole region was found to be accessible in the near future at the LHC. It was also argued that the coannihilation region has been already excluded by the LHC experiments. Thus, dilutions are required to take place in early universe, since otherwise the relic neutralino overcloses the universe, or the R-parity might be violated slightly.

\section*{Acknowledgments}
This work was supported by Grand-in-Aid for Scientific research from
the Ministry of Education, Science, Sports, and Culture (MEXT), Japan,
No. 23740172 (M.E.), No. 21740164 (K.H.), No. 22244021 (K.H.) and No. 22-7585 (N.Y.).
S.I. is supported by JSPS Grant-in-Aid for JSPS Fellows.
This work was supported by World Premier International Research Center Initiative (WPI Initiative), MEXT, Japan.

%%%%%%%%%%%%%%%%%%%%%%%%%%%%%%%%%%%%%%%%%%%%

\end{document}